\documentclass[3p,times]{elsarticle}
\usepackage{ecrc}
\usepackage{epstopdf}
\def\citet{\cite}
\newcommand{\lyxdot}{.}

\volume{00}
\firstpage{1}
\journalname{Nuclear Physics A}
\runauth{K. Werner et al.}
\jid{nupha}
\jnltitlelogo{Nuclear Physics A}
\usepackage{graphicx}
\usepackage{amsmath,amssymb}
\begin{document}
\begin{frontmatter}
\title{A unified description of the reaction dynamics from pp to pA to AA collisions}
\author[1]{ K.$\,$Werner    }
\author[1]{ B. Guiot        }
\author[2,3]{ Iu.$\,$Karpenko }
\author[4]{ T.$\,$Pierog    }
\address[1]{ SUBATECH, University of Nantes -- IN2P3/CNRS-- EMN, Nantes, France  }
\address[2]{ Bogolyubov Institute for Theoretical Physics, Kiev 143, 03680, Ukraine  }
\address[3]{ FIAS, Johann Wolfgang Goethe Universitaet, Frankfurt am Main, Germany  }
\address[4]{ Karlsruhe Inst. of Technology, KIT, Campus North, Inst. f. Kernphysik, Germany  }

\begin{abstract}

There is little doubt that in heavy ion collisions at the LHC and
RHIC, we observe a hydrodynamically expanding system, providing strong
evidence for the formation of a Quark Gluon Plasma (QGP) in the early
stage of such collisions. These observations are mainly based on results
on azimuthal anisotropies, but also on particle spectra of identified
particles, perfectly compatible with a hydrodynamic evolution. Surprisingly,
in p-Pb collisions one observes a very similar behavior, and to some
extent even in p-p. We take these experimental observations as a strong
support for a unified approach to describe proton-proton (p-p), proton-nucleus
(p-A), and nucleus-nucleus (A-A) collisions, with a plasma formation
even in tiny systems as in p-p scatterings. 

\end{abstract}

\begin{keyword}
flow \sep QGP \sep EPOS3

\end{keyword}

\end{frontmatter}

\section{Experimental evidence for a unified picture }

Collective hydrodynamic flow seems to be well established in heavy
ion (HI) collisions at energies between 200 and 2760 AGeV, whereas
p-p and p-A collisions are often considered to be simple reference
systems, showing {}``normal'' behavior, such that deviations of
HI results with respect to p-p or p-A reveal {}``new physics''.
Surprisingly, the first results from p-Pb at 5 TeV on the transverse
momentum dependence of azimuthal anisotropies and particle yields
are very similar to the observations in HI scattering \citet{cms,alice}. 

Information about flow can be obtained via studying two particle correlations
as a function of the pseudorapidity difference $\Delta\eta$ and the
azimuthal angle difference $\Delta\phi$. So-called ridge structures
(at $\Delta\phi=0$, very broad in $\Delta\eta$) have been observed
first in heavy ion collisions, later also in pp \citet{cms10} and
very recently in p-Pb collisions \citet{alice12,cms12,atlas13}, as
shown in fig. \ref{fig:flow-1}. In case of heavy ions, these structures
appear naturally in models employing a hydrodynamic expansion, in
an event-by-event treatment -- provided the azimuthal asymmetries
are (essentially) longitudinally invariant, as in the string model
approach. 

To clearly pin down the origin of such structures in small systems,
one needs to consider identified particles. In the fluid dynamical
scenario, where particles are produced in the local rest frame of
fluid cells characterized by transverse velocities, large mass particles
(compared to low mass ones) are pushed to higher transverse momenta.
When plotting ratios {}``heavy over light'' versus $p_{t}$, one
observes a peak at intermediate $p_{t}$, more and more pronounced
with increasing flow. In fig. \ref{fig:flow-2}, we show lambda over
kaon ratios from ALICE for Pb-Pb (right plot, \citet{alice13}) and
p-Pb (left plot, \citet{alice}). In both cases, for more central
collisions the peak is more pronounced, which may be interpreted as
stronger radial flow compared to more peripheral collisions. Unexpectedly,
the p-Pb results are qualitatively very similar to the Pb-Pb ones.
\begin{figure}[tb]
\centering{}%
\end{figure}
\begin{figure}[tb]
\begin{centering}
~
\par\end{centering}

\begin{centering}
\hspace*{-0.4cm}\includegraphics[scale=0.39]{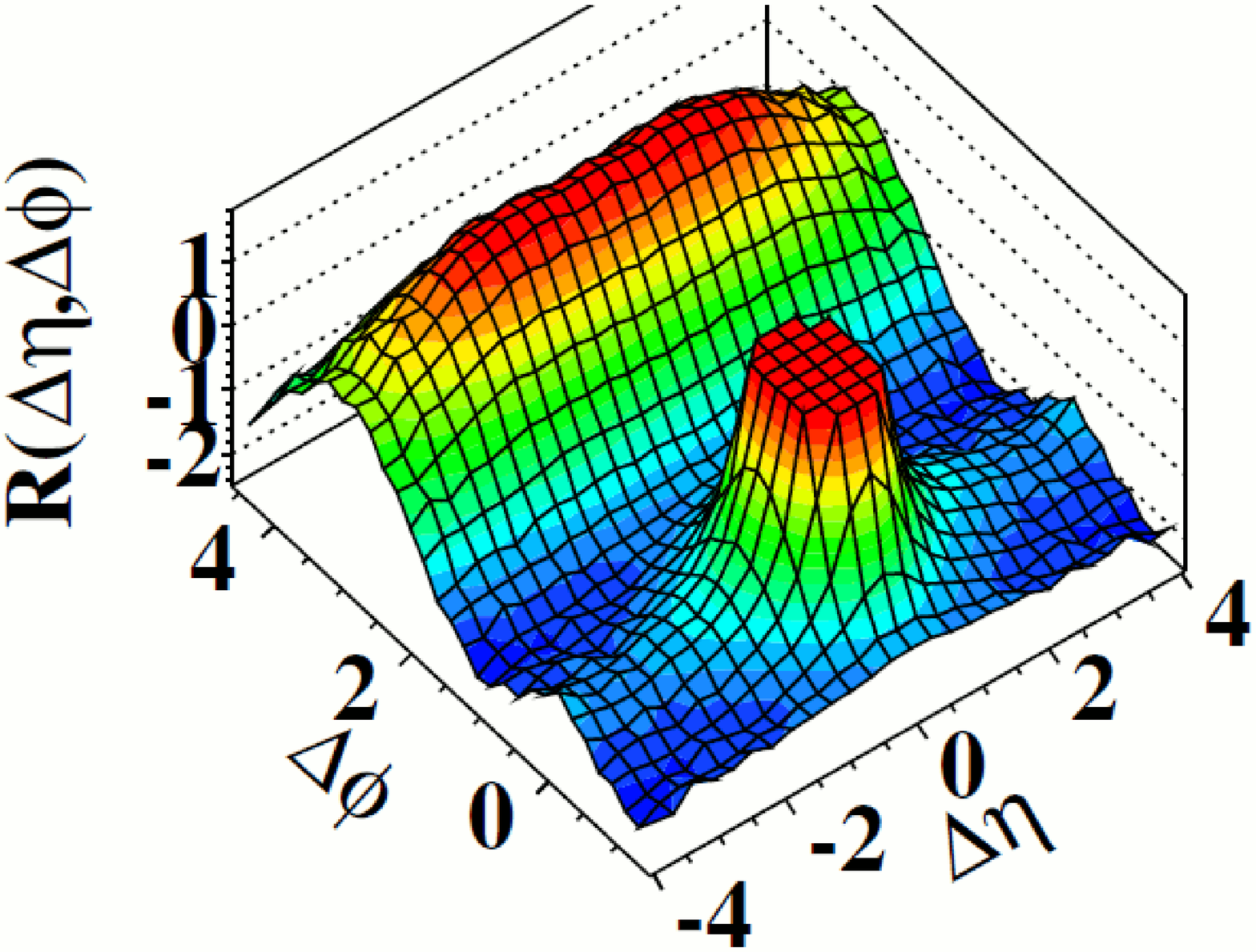}$\quad$\includegraphics[scale=0.43]{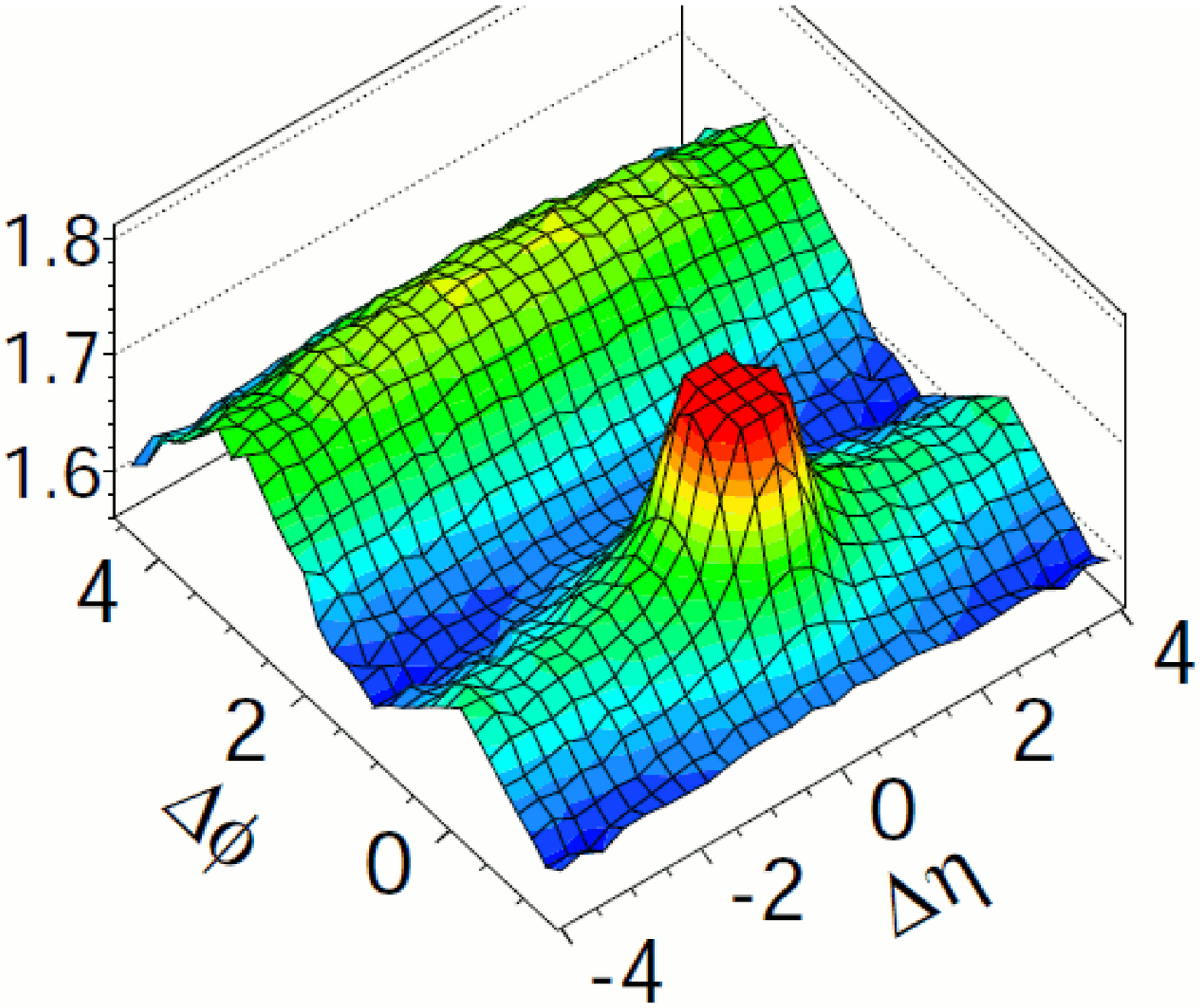}
\par\end{centering}

\caption{(Color online) Two particle correlation functions as a function of
the pseudorapidity difference $\Delta\eta$ and the azimuthal angle
difference $\Delta\phi$, from the CMS experiment. Left: p-p, right:
p-Pb. \label{fig:flow-1} In both cases, the jet peak at $\Delta\eta$=0
and $\Delta\phi=0$ has been truncated, for better visibility. In
both cases a {}``ridge structure'' shows up, at $\Delta\phi=0$
and very broad in $\Delta\eta$.}

\vspace{0.5cm}

\begin{centering}
\includegraphics[scale=0.55]{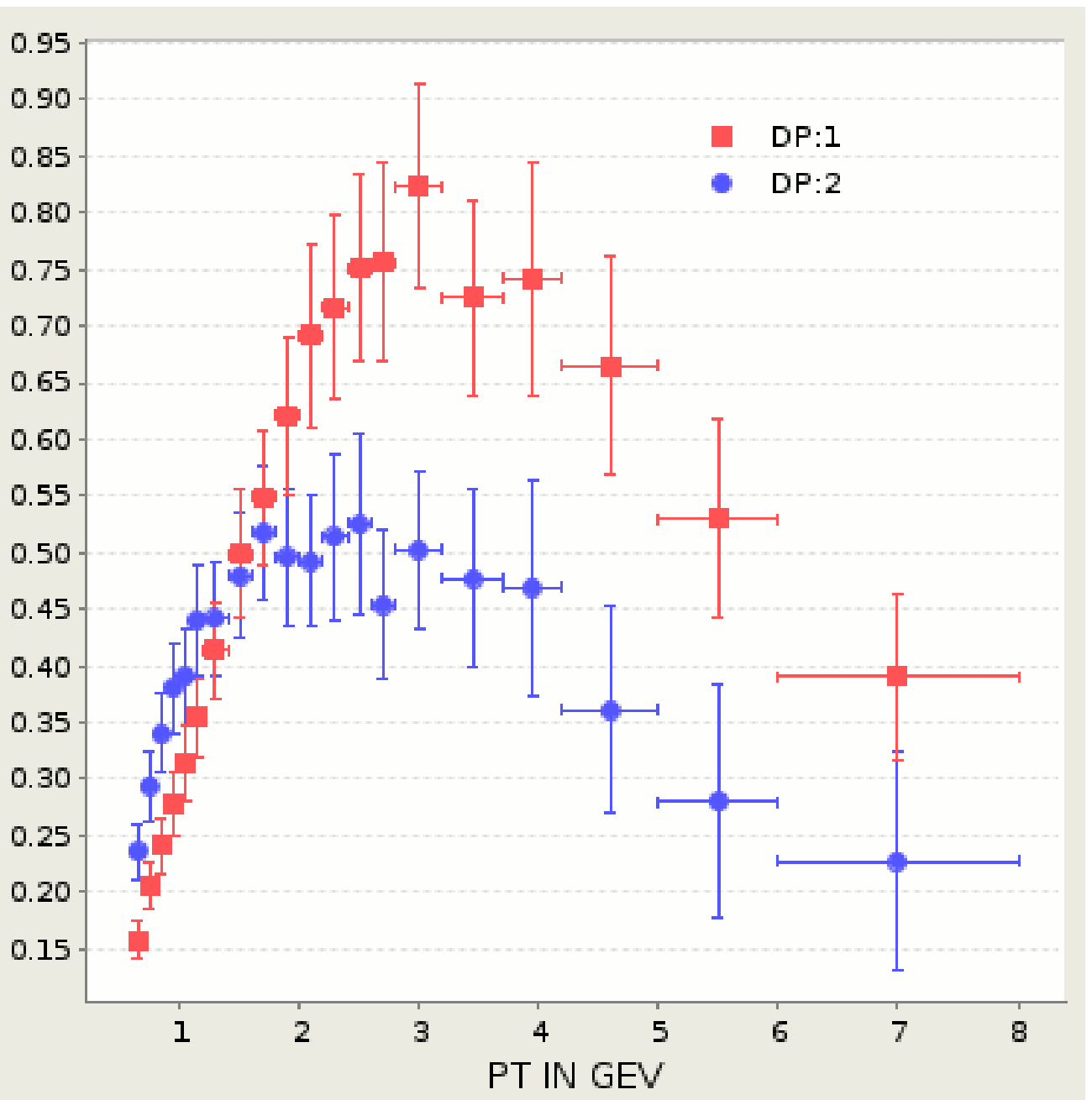}$\qquad$\includegraphics[scale=0.55]{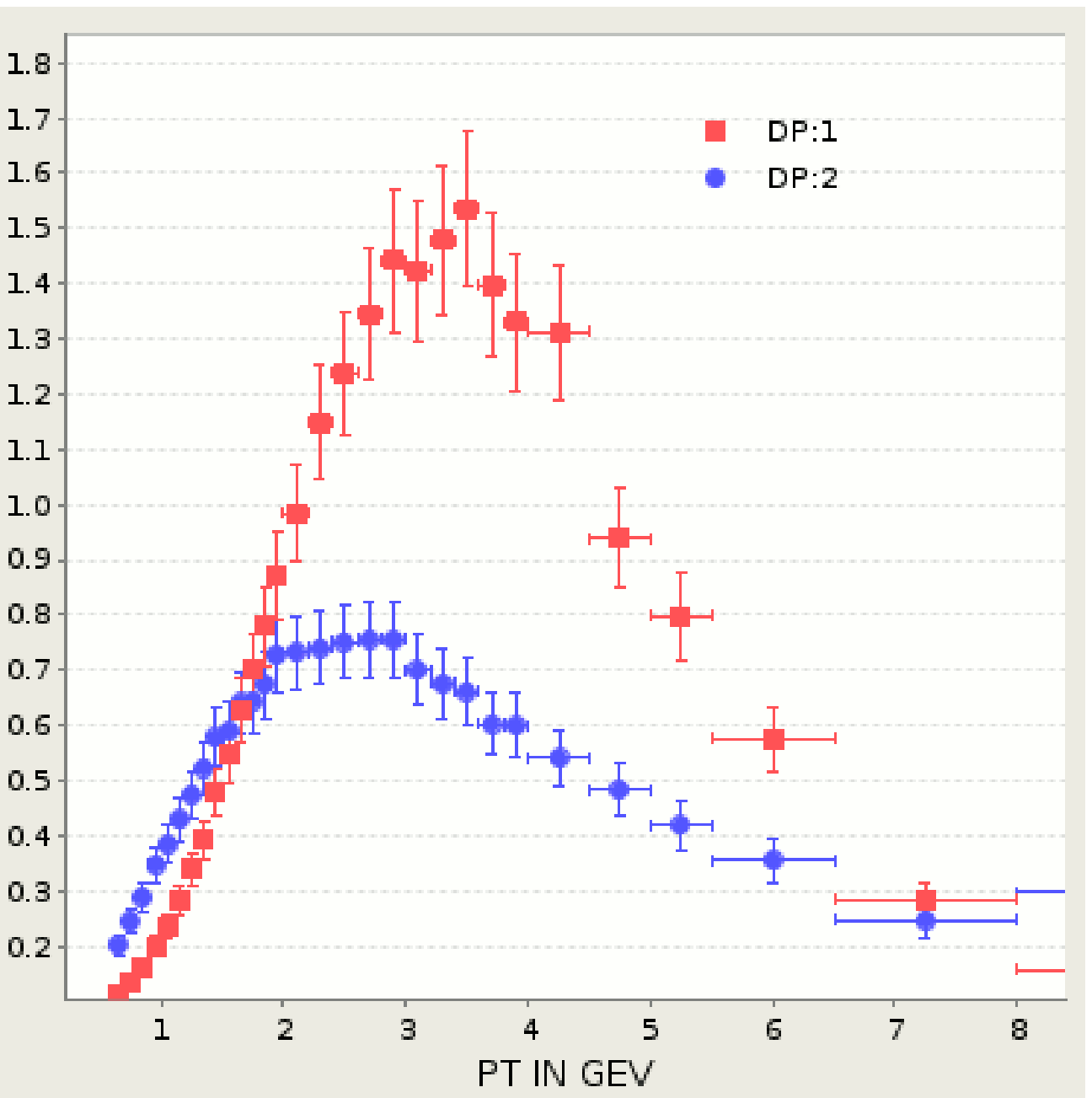}
\par\end{centering}

\begin{centering}
~
\par\end{centering}

\caption{(Color online) Lambda over kaon ratios versus $p_{t}$ from ALICE
for p-Pb (left) and Pb-Pb (right). In both cases two centrality classes
are shown: 0-5\% (red) and 60-80\% (blue).\label{fig:flow-2} The
more pronounced peaks in more central collisions (in both cases) may
be interpreted as stronger radial flow. }
\end{figure}

\begin{figure}[tb]
\begin{centering}
\includegraphics[scale=0.28]{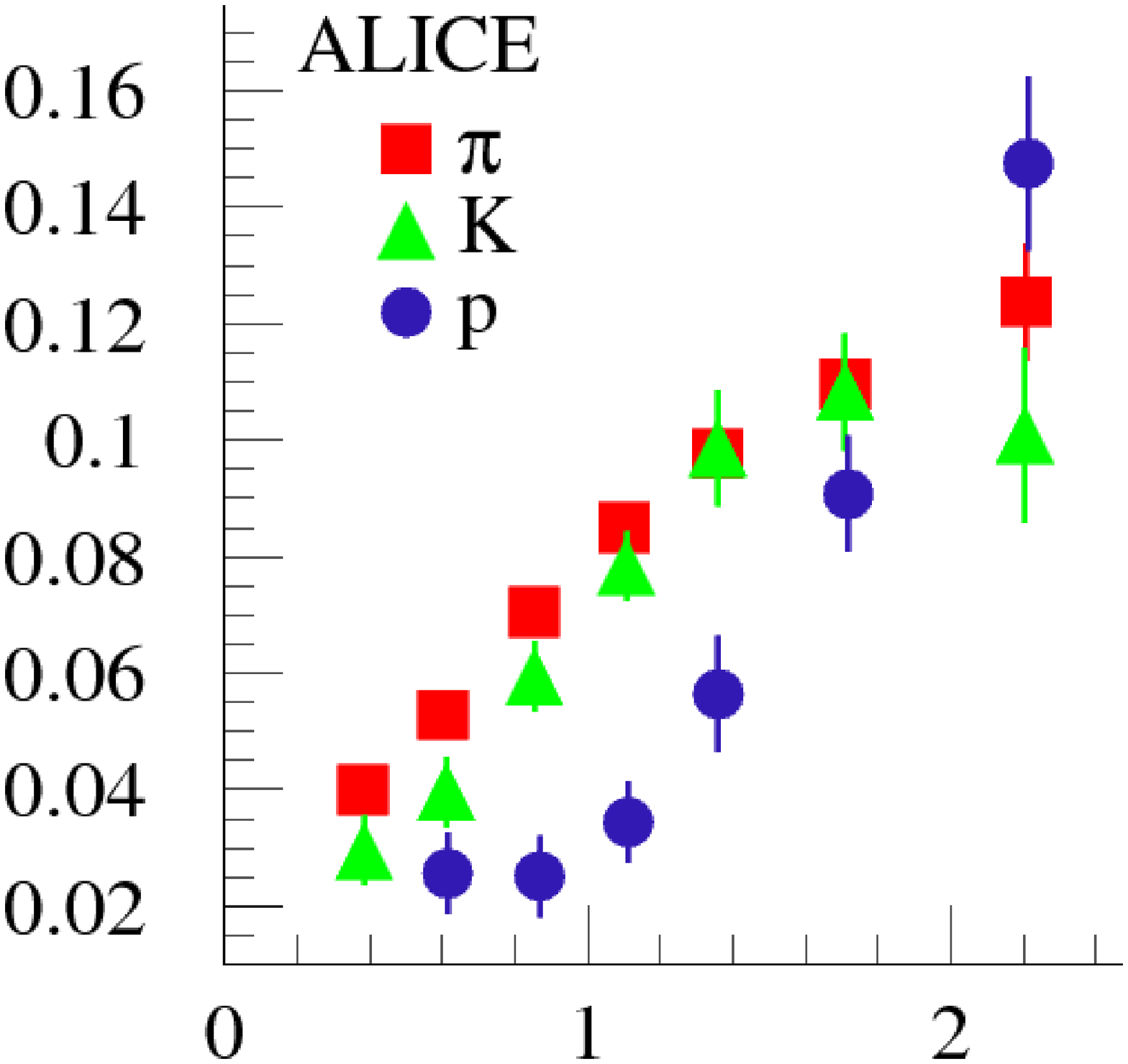}$\qquad\qquad$ \includegraphics[scale=0.44]{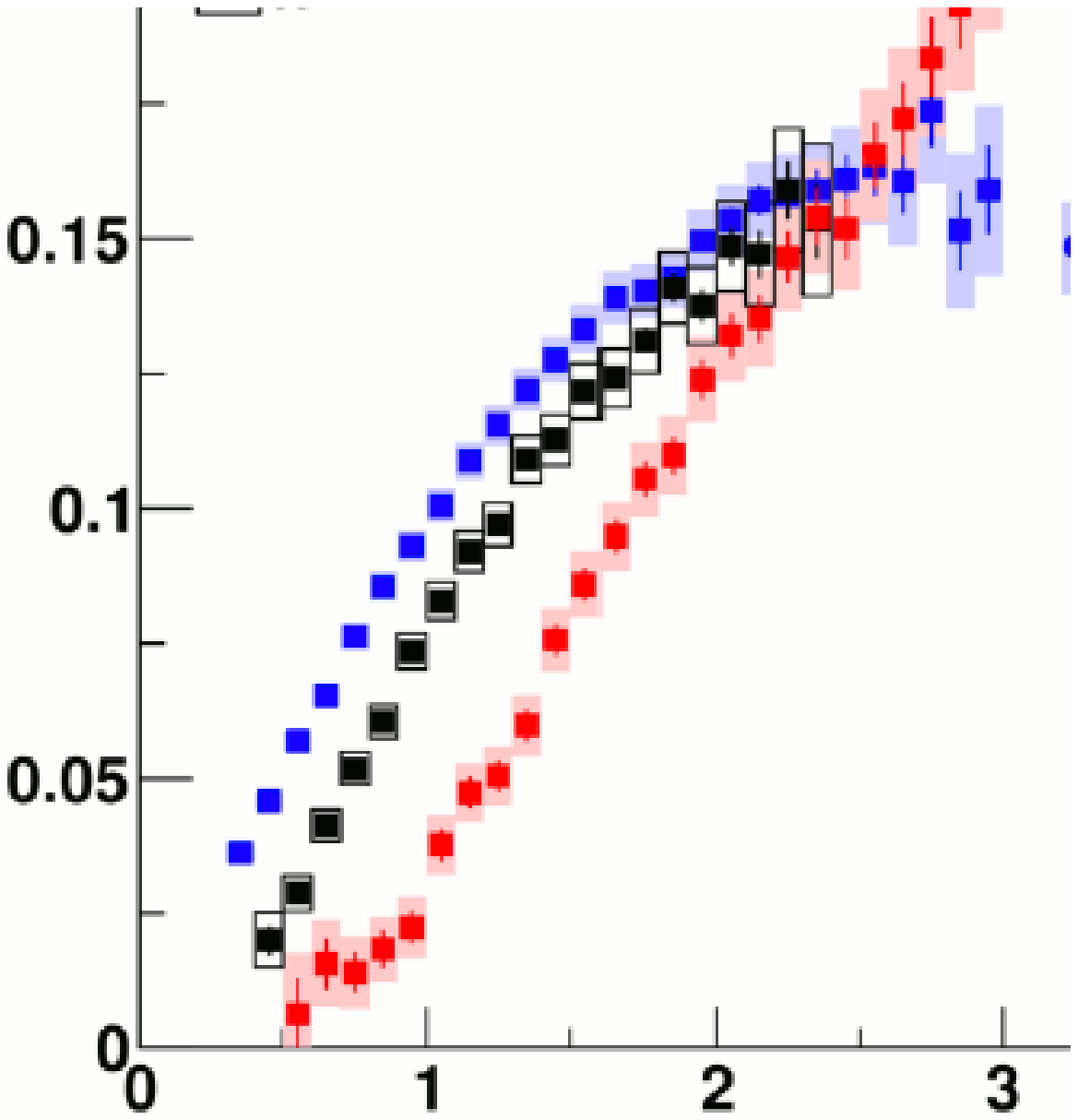}
\par\end{centering}

\caption{(Color online) Elliptical flow coefficient $v_{2}$ as a function
of $p_{t}$ from ALICE for p-Pb (left) and Pb-Pb (right). We show
results for pions, kaons, and protons.\label{fig:flow-3}}
\end{figure}

Finally, one can combine the power of dihadron correlations and particle
identification: The mass effect discussed above for particle spectra,
is also very clearly visible in correlations, leading to the so-called
mass-splitting in the elliptical flow coefficient $v_{2}$ as a function
of $p_{t}$, as shown in fig. \ref{fig:flow-3}, where we plot $v_{2}$
for different hadrons for Pb-Pb (right plot, \citet{alice12b}) and
p-Pb (left plot, \citet{alice13b}). In both cases, one can clearly
see the separation of particles of different masses. Again, the p-Pb
results are very similar to the Pb-Pb ones. %

There are many more examples, where p-Pb (and even p-p) shows a very
similar behavior compared to Pb-Pb, which strongly supports the idea
of a unified picture in all these different reactions, from p-p to
Pb-Pb.

\section{Unified description of the reaction dynamics}

We take the experimental observations discussed in the previous section
as an excellent motivation and justification for a unified description
of the dynamics of ALL reactions, from p-p to AA. In this picture,
the same procedure applies, referred to as EPOS3 \citet{epos3}, based
on several stages:
\begin{description}
\item [{\underbar{Initial~conditions.}}] A Gribov-Regge multiple scattering
approach is employed \citet{hajo}, where the elementary object (by
definition called Pomeron) is a DGLAP parton ladder, using in addition
a CGC motivated saturation scale \citet{sat1} for each Pomeron, of
the form\textbf{ $Q_{s}\propto N_{\mathrm{part}}\,\hat{s}^{\lambda}$},
where $N_{\mathrm{part}}$ is the number of nucleons connected the
Pomeron in question, and $\hat{s}$ its energy. The parton ladders
are treated as classical relativistic (kinky) strings.
\item [{\underbar{Core-corona~approach.}}] At some early proper time $\tau_{0}$,
one separates fluid (core) and escaping hadrons, including jet hadrons
(corona), based on the momenta and the density of string segments
\citet{core,epos3}. The corresponding energy-momentum tensor of the
core part is transformed into an equilibrium one, needed to start
the hydrodynamical evolution. This is based on the hypothesis that
equilibration happens rapidly and affects essentially the space components
of the energy-momentum tensor.
\item [{\underbar{Viscous~hydrodynamic~expansion.}}] Starting from the
initial proper time $\tau_{0}$, the core part of the system evolves
according to the equations of relativistic viscous hydrodynamics \citet{epos3,yuri},
where we use presently $\eta/s=0.08$. A cross-over equation-of-state
is used, compatible with lattice QCD \citet{lattice,kw1}. 
\item [{\underbar{Statistical~hadronization}}] The {}``core-matter''
hadronizes on some hypersurface defined by a constant temperature
$T_{H}$, where a so-called Cooper-Frye procedure is employed, using
equilibrium hadron distributions, see \citet{kw1}.
\item [{\underbar{Final~state~hadronic~cascade}}] After hadronization,
the hadron density is still big enough to allow hadron-hadron rescatterings.
For this purpose, we use the UrQMD model \citet{urqmd}.
\end{description}
The above procedure is employed for each event (event-by-event procedure).

Whereas our approach is described in detail in \citet{epos3}, referring
to older works \citet{hajo,core,kw1}, we confine ourselves here to
a couple of remarks, to selected items. The initial conditions are
generated in the Gribov-Regge multiple scattering framework. Our formalism
is referred to as {}``Parton-Based Gribov-Regge Theory'' (PBGRT)
and described in very detail in \citet{hajo}, see also \citet{epos3}
for all the details of the present (EPOS3) implementation. The fundamental
assumption of the approach is the hypothesis that the S-matrix is
given as a product of elementary objects, referred to as Pomerons.
Once the Pomeron is specified (taken as a DGLAP parton ladder, including
a saturation scale), everything is completely determined. Employing
cutting rule techniques, one may %
\begin{figure}[tb]
\begin{minipage}[t]{1\columnwidth}%
\noindent \begin{center}
\begin{minipage}[c]{0.35\columnwidth}%
\noindent \begin{center}
{\Large \[
\sigma^{\mathrm{tot}}\;=\quad\sum_{\mathrm{cut\, P}}\int\quad\sum_{\mathrm{uncut\, P}}\int\]
}
\par\end{center}%
\end{minipage}%
\begin{minipage}[c]{0.4\columnwidth}%
\noindent \begin{center}
\includegraphics[scale=0.25]{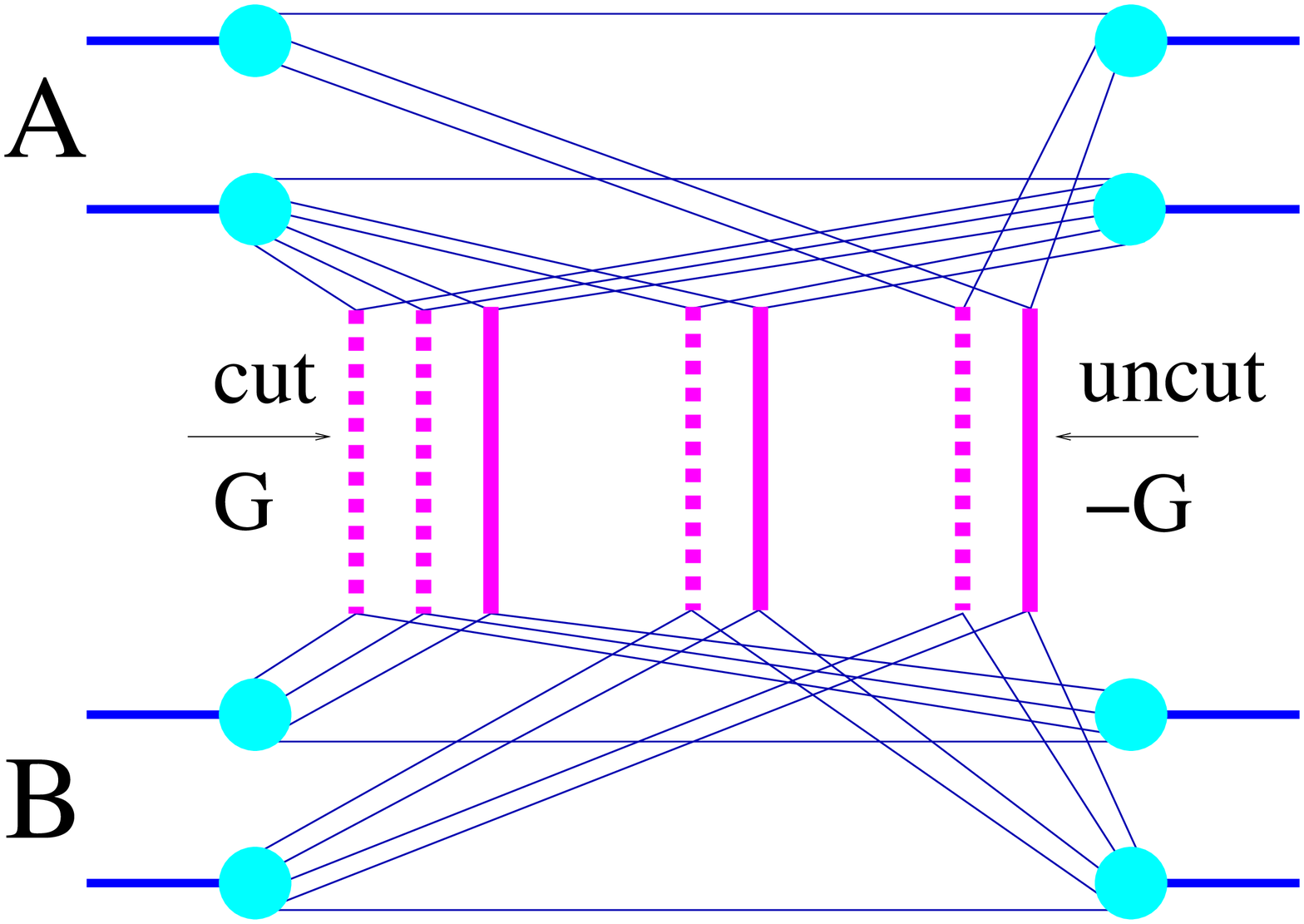}
\par\end{center}%
\end{minipage}\vspace{-0.6cm}

\par\end{center}

\noindent \begin{center}
\begin{minipage}[t]{0.95\columnwidth}%
{\LARGE \[
\qquad\qquad\quad\qquad\underbrace{\qquad\qquad\qquad\quad\qquad\qquad\qquad\qquad}\]
}\vspace{-0.6cm}
\end{minipage}
\par\end{center}

\begin{minipage}[t]{0.95\columnwidth}%
{\Large \[
\qquad\qquad\qquad\qquad\qquad d\sigma_{\mathrm{exclusive}}\]
~}%
\end{minipage}

~%
\end{minipage}

\caption{(Color online) PBGRT formalism: The total cross section expressed
in terms of cut (dashed lines) and uncut (solid lines) Pomerons, for
nucleus-nucleus, proton-nucleus, and proton-proton collisions. Partial
summations allow to obtain exclusive cross sections, the mathematical
formulas can be found in \citet{hajo}, or in a somewhat simplified
form in \citet{epos3}. \label{muscatt}}
\end{figure}
express the total cross section in terms of cut and uncut Pomerons,
as sketched in fig. \ref{muscatt}. The great advantage of this approach:
doing partial summations, one obtains expressions for partial cross
sections $d\sigma_{\mathrm{exclusive}}$, for particular multiple
scattering configurations, based on which the Monte Carlo generation
of configurations can be done. No additional approximations are needed.
The above multiple scattering picture is used for p-p, p-A, and A-A. 

\begin{figure}[tb]
\begin{centering}
\includegraphics[scale=0.65]{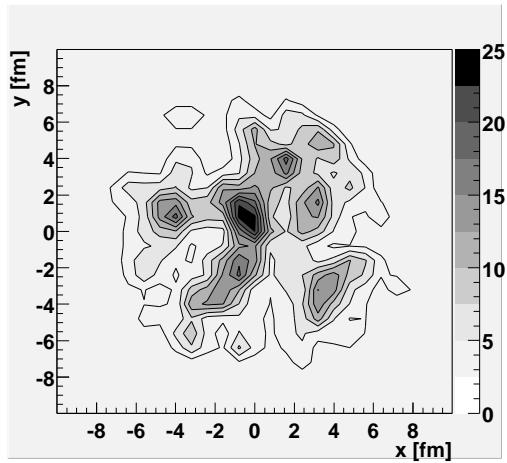}
\par\end{centering}

\caption{(Color online) Initial condition for QGP as obtained in the PBGRT
framework in the year 2000, see ref \citet{bumpy}. \label{fig:bumpy}}
\end{figure}

Based on the PBGRT approach, we obtain in A-A collisions a very large
number of strings, but the randomness of their transverse positions
leads to {}``bumpy'' energy density distributions in the transverse
plane at $\tau_{0}$, as published for the first time in the year
2000, see fig. 21 of \citet{bumpy}, reproduced as fig. \ref{fig:bumpy}
in this paper.

\begin{figure}[tb]
\begin{centering}
\includegraphics[angle=270,scale=0.28]{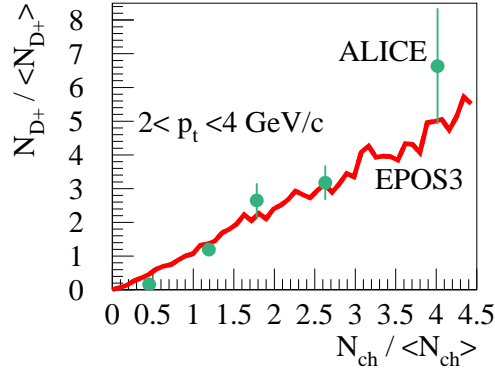}
\par\end{centering}

\caption{(Color online) $D$ meson multiplicity versus charged particle multiplicity
(both properly normalized). We show EPOS3 simulations as red line,
and preliminary data from ALICE \citet{zaida}. \label{fig:dmesons}}
\end{figure}

Our multiple scattering approach leads in a natural way to very simple
features when it comes to relating soft and hard particle production.
Be $N_{\mathrm{hard}}$ the multiplicity of some {}``hard'' particle
production (like the $D$ meson multiplicity) and $N_{\mathrm{ch}}$
the usual charged particle multiplicity in some phase space interval.
We expect to first approximation a linear relation, $N_{\mathrm{hard}}\propto N_{\mathrm{ch}},$
since both are proportional to the number $N_{\mathrm{Pom}}$ of Pomerons.
We obtain indeed such a linear behavior, as seen in fig. \ref{fig:dmesons}.

\begin{figure}[tb]
\noindent \begin{centering}
\begin{minipage}[c]{0.35\columnwidth}%
\includegraphics[angle=270,scale=0.28]{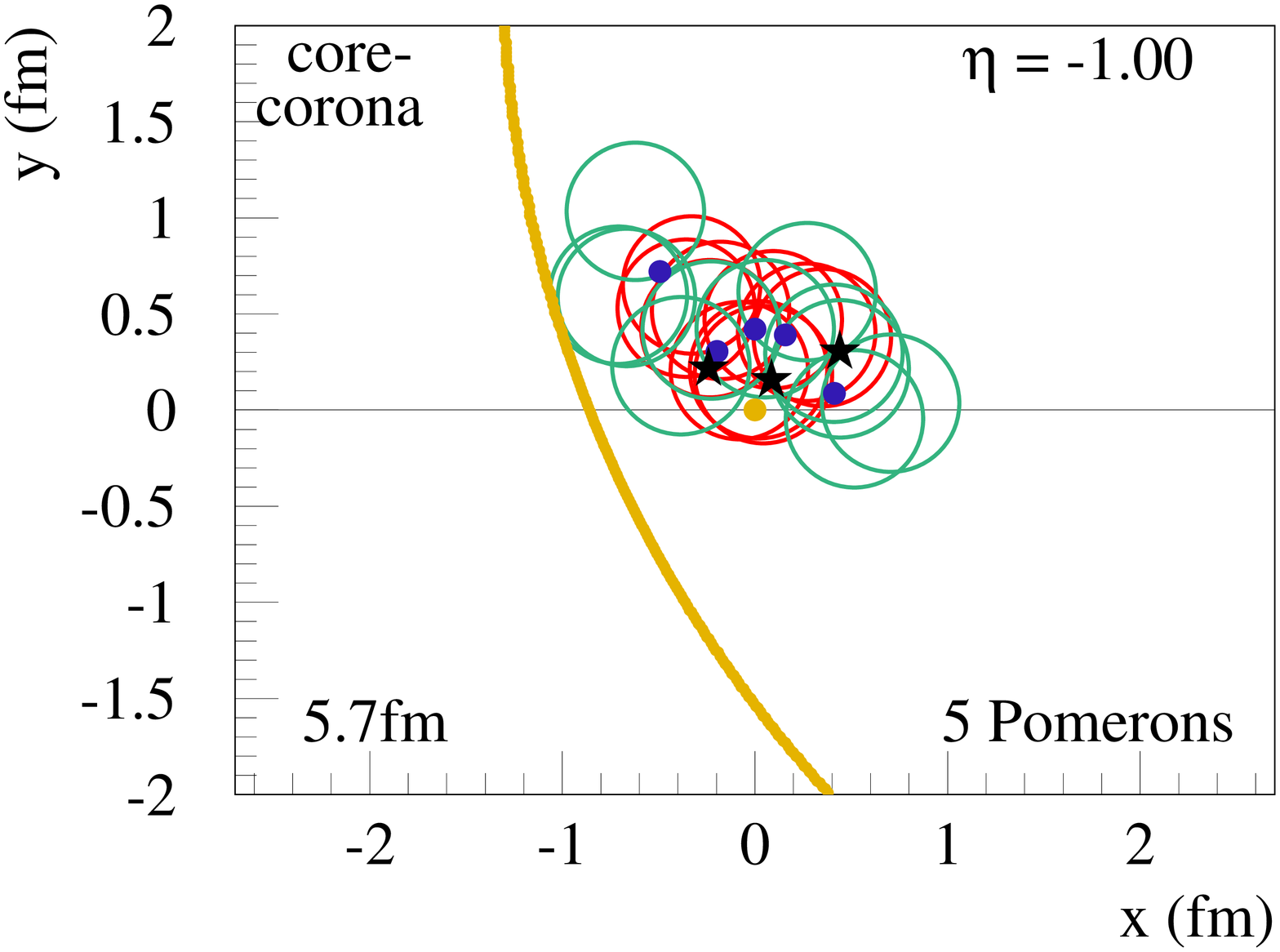}%
\end{minipage}%
\begin{minipage}[c]{0.1\columnwidth}%
~{\Large p-Pb}{\Large \par}

~\vspace*{2cm}

~%
\end{minipage}$\qquad$%
\begin{minipage}[c]{0.4\columnwidth}%
\includegraphics[angle=270,scale=0.28]{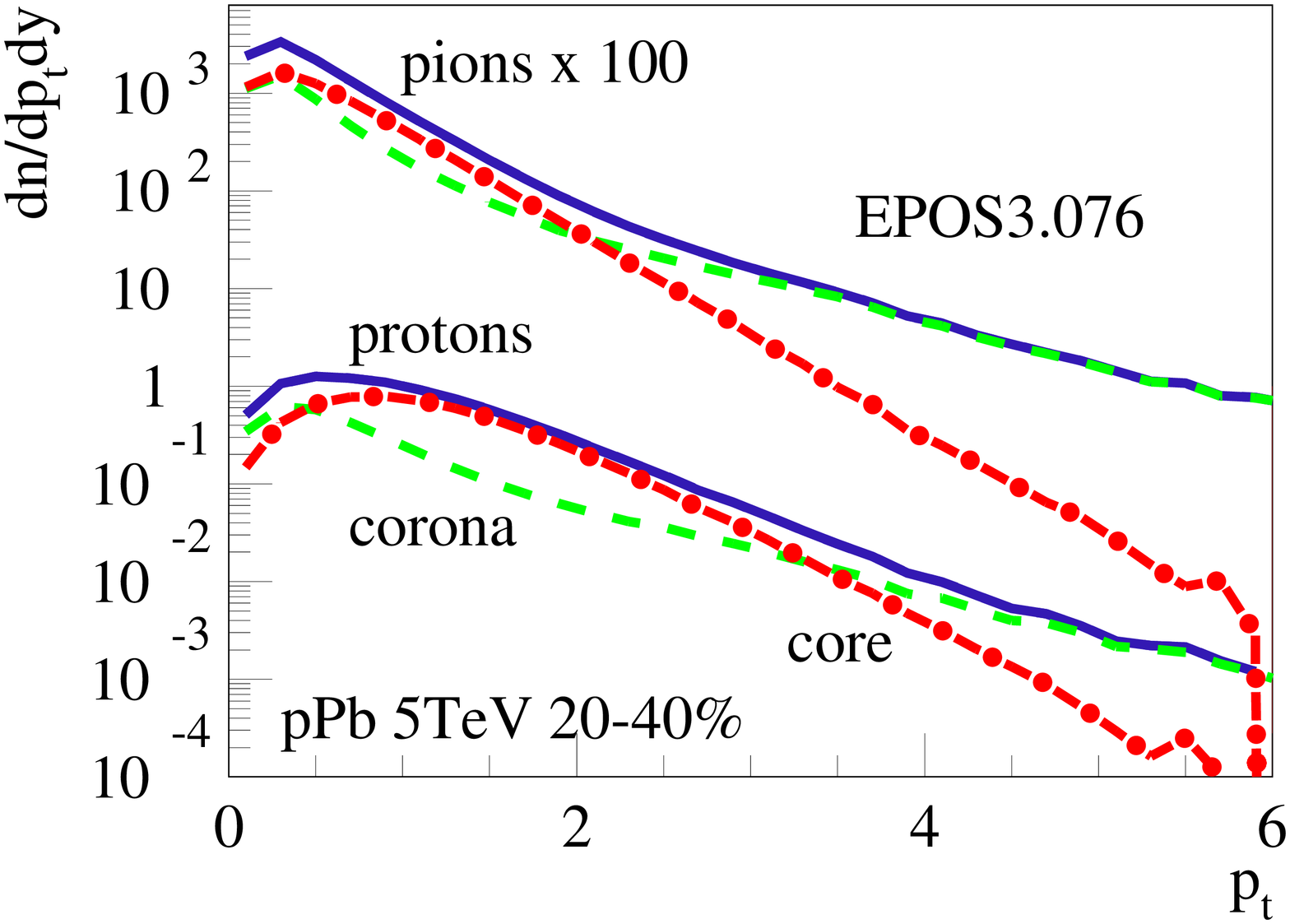}%
\end{minipage}
\par\end{centering}

\caption{(Color online) Left: Core-corona separation in a semi-peripheral p-Pb
collision. Right: the corresponding core and corona contributions
to the $p_{t}$ spectra of pions and protons.\label{fig:core-example}}
\end{figure}

To understand the results later in this paper, we will discuss an
example of core-corona separation in a semi-peripheral p-Pb collision,
as shown in fig. \ref{fig:core-example}. Shown (in the left figure)
are string %
segments in the transverse plane, red (core) and green (corona) ones.
There are sufficient overlapping core string segments to provide a
core of plasma matter, showing a (short) hydrodynamic expansion, quickly
building up flow. In the right figure, we plot the contribution from
core and corona to the $p_{t}$ spectra of pions and protons. In particular
for protons, the core dominates at intermediate $p_{t}$ (mass effect,
as discussed earlier).

\section{Comparison with data and other models}

The unified approach discussed above gives quite good results concerning
heavy ion collisions -- as many other models do as well. Therefore
we concentrate in this paper p-Pb scattering.

In the following, we will compare experimental data on identified
particle production with our simulation results (referred to as EPOS3),
and in addition to some other models, as there are QGSJETII \citet{qgsjet},
AMPT \citet{ampt}, and EPOS$\,$LHC \citet{eposlhc}. The QGSJETII
model is also based on Gribov-Regge multiple scattering, but there
is no fluid component. The main ingredients of the AMPT model are
a partonic cascade and then a hadronic cascade, providing in this
way some {}``collectivity''. EPOS$\,$LHC is a tune (using LHC data)
of EPOS1.99. As all EPOS1 models, it contains flow, put in by hand,
parametrizing the collective flow at freeze-out. Finally, the approach
discussed in this paper (EPOS3) contains a full viscous hydrodynamical
simulation. So it is interesting to compare these four models, since
they differ considerably concerning the implementation of flow, from
full hydrodynamical flow in EPOS3 to no flow in QGSJETII.

\begin{figure}[tb]
\begin{minipage}[c]{1\columnwidth}%
\begin{center}
\vspace*{-0.4cm}
\par\end{center}

\begin{center}
\includegraphics[angle=270,scale=0.28]{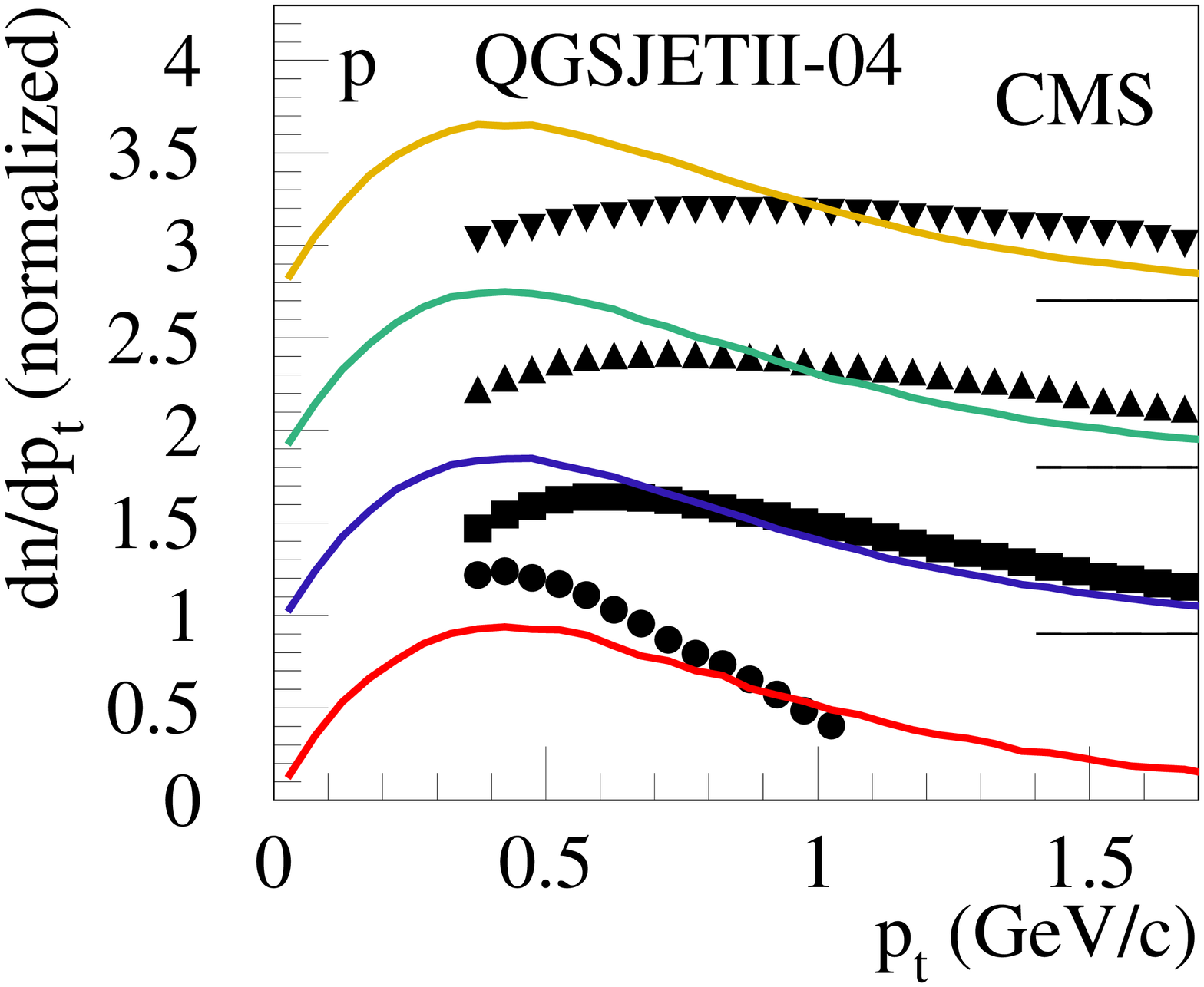}$\qquad$\includegraphics[angle=270,scale=0.28]{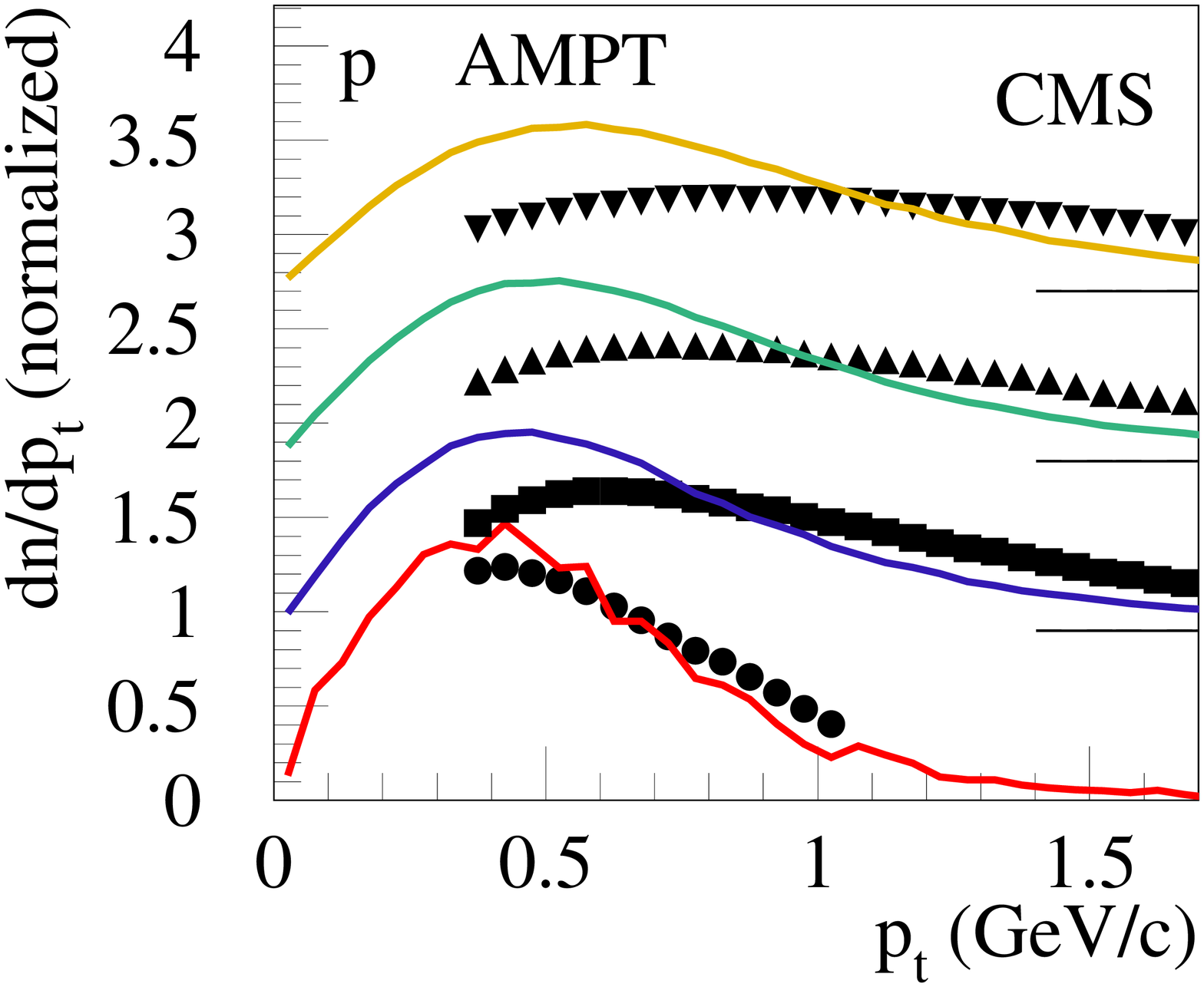}\vspace*{-0.6cm}
\par\end{center}

\begin{center}
\includegraphics[angle=270,scale=0.28]{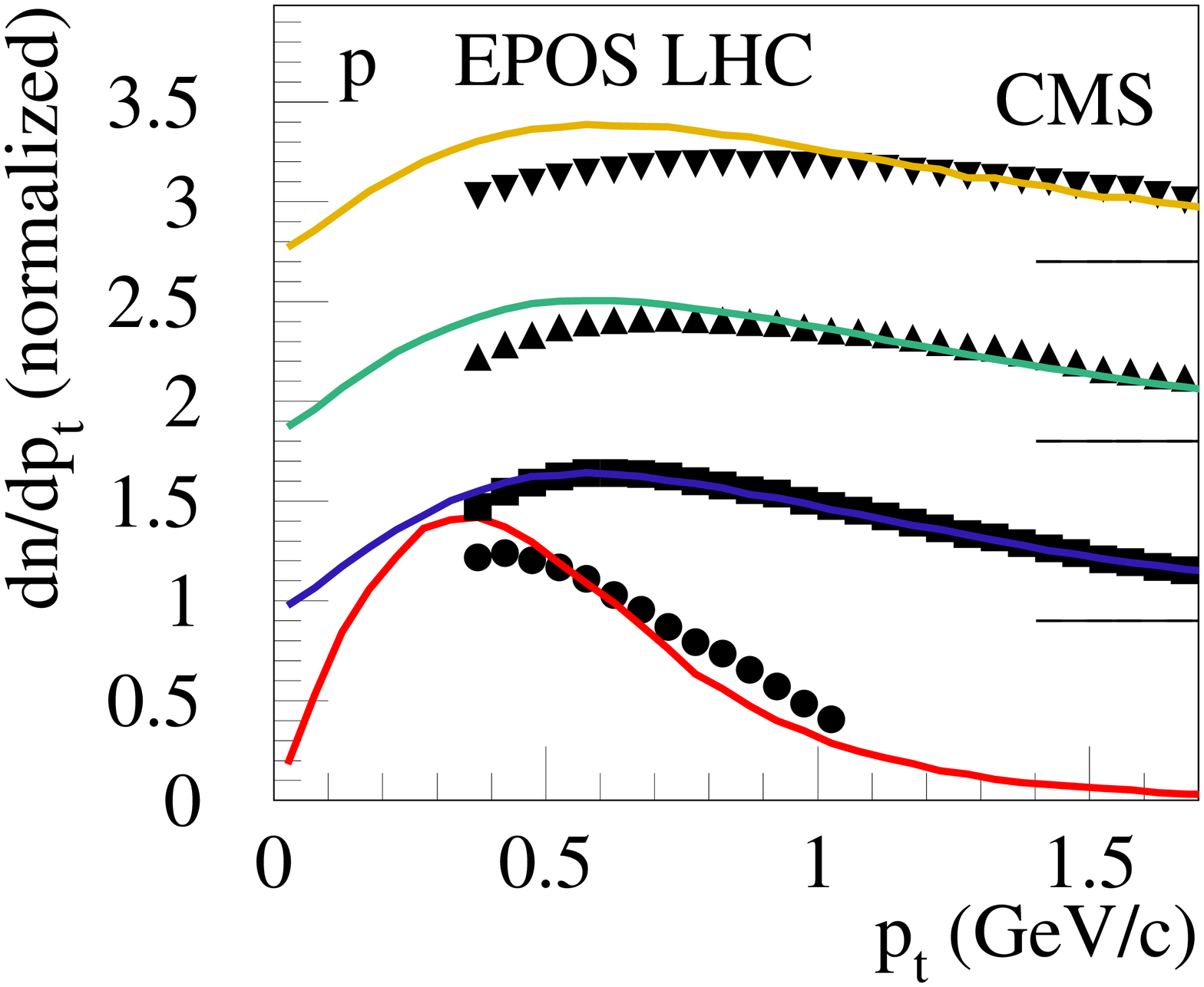}$\qquad$\includegraphics[angle=270,scale=0.28]{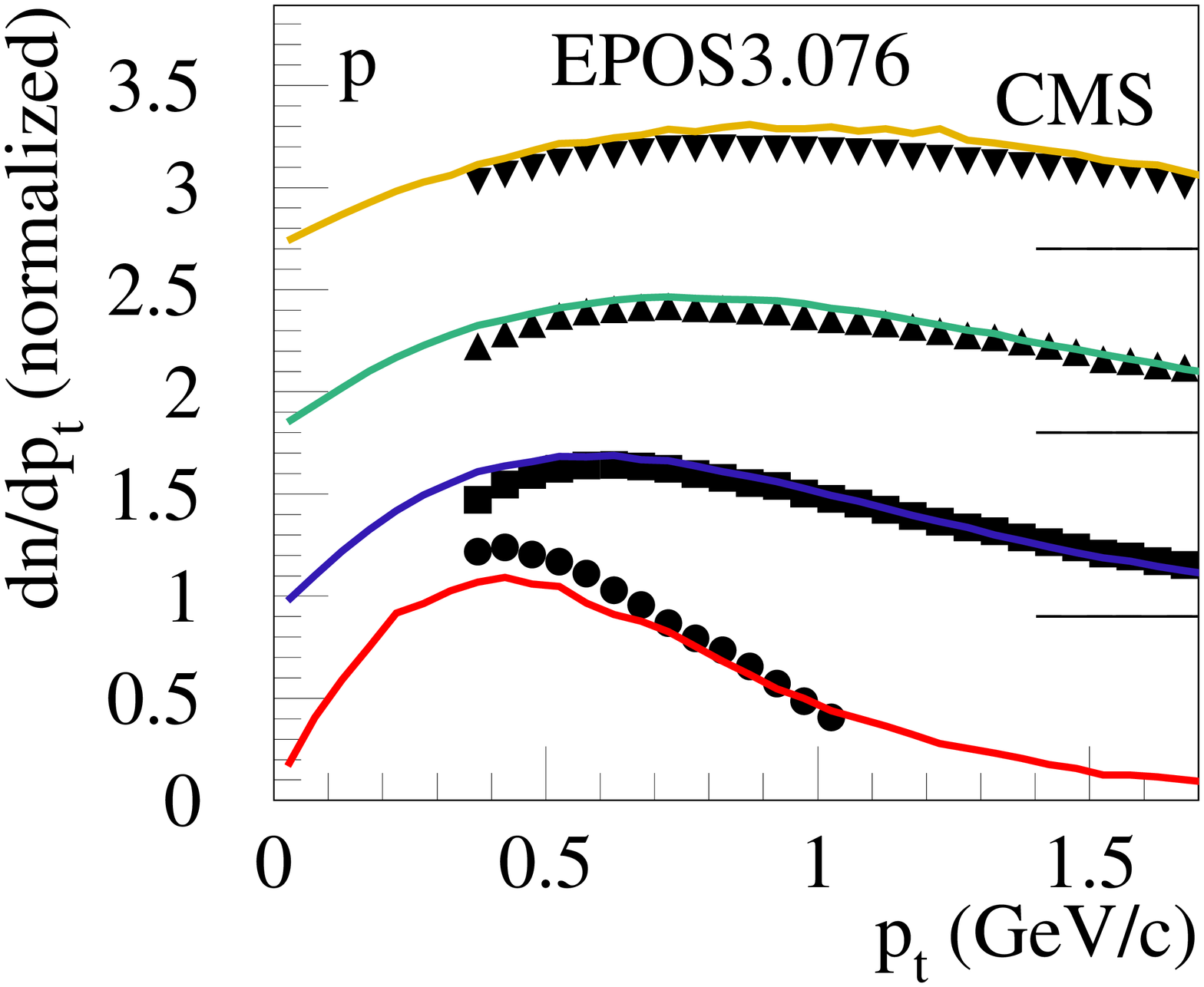}
\par\end{center}%
\end{minipage}

\noindent \caption{(Color online) Transverse momentum spectra of protons in p-Pb scattering
at 5.02 TeV, for four different multiplicity classes with mean values
(from bottom to top) of 8, 84, 160, and 235 charged tracks. We show
data from CMS \citet{cms} (symbols) and simulations from QGSJETII,
AMPT, EPOS$\,$LHC, and EPOS3, as indicated in the figures. \label{fig:cms1}}
\end{figure}
The CMS collaboration published a detailed study \citet{cms} of the
multiplicity dependence of (normalized) transverse momentum spectra
in p-Pb scattering at 5.02 TeV. The multiplicity (referred to as $N_{\mathrm{track}}$)
counts the number of charged particles in the range $|\eta|<2.4$.
In fig. \ref{fig:cms1}, we compare experimental data \citet{cms}
for protons (black symbols) with the simulations from QGSJETII (upper
left figure), AMPT (upper right), EPOS$\,$LHC (lower left), and EPOS3
(lower right). The different curves in each figure refer to different
multiplicities, with mean values (from bottom to top) of 8, 84, 160,
and 235 charged tracks. They are shifted relative to each other by
a constant amount. %
\begin{figure}[tb]
\begin{raggedleft}
\begin{minipage}[c]{1\columnwidth}%
\begin{center}
\vspace*{-0.4cm}
\par\end{center}

\begin{center}
\includegraphics[angle=270,scale=0.28]{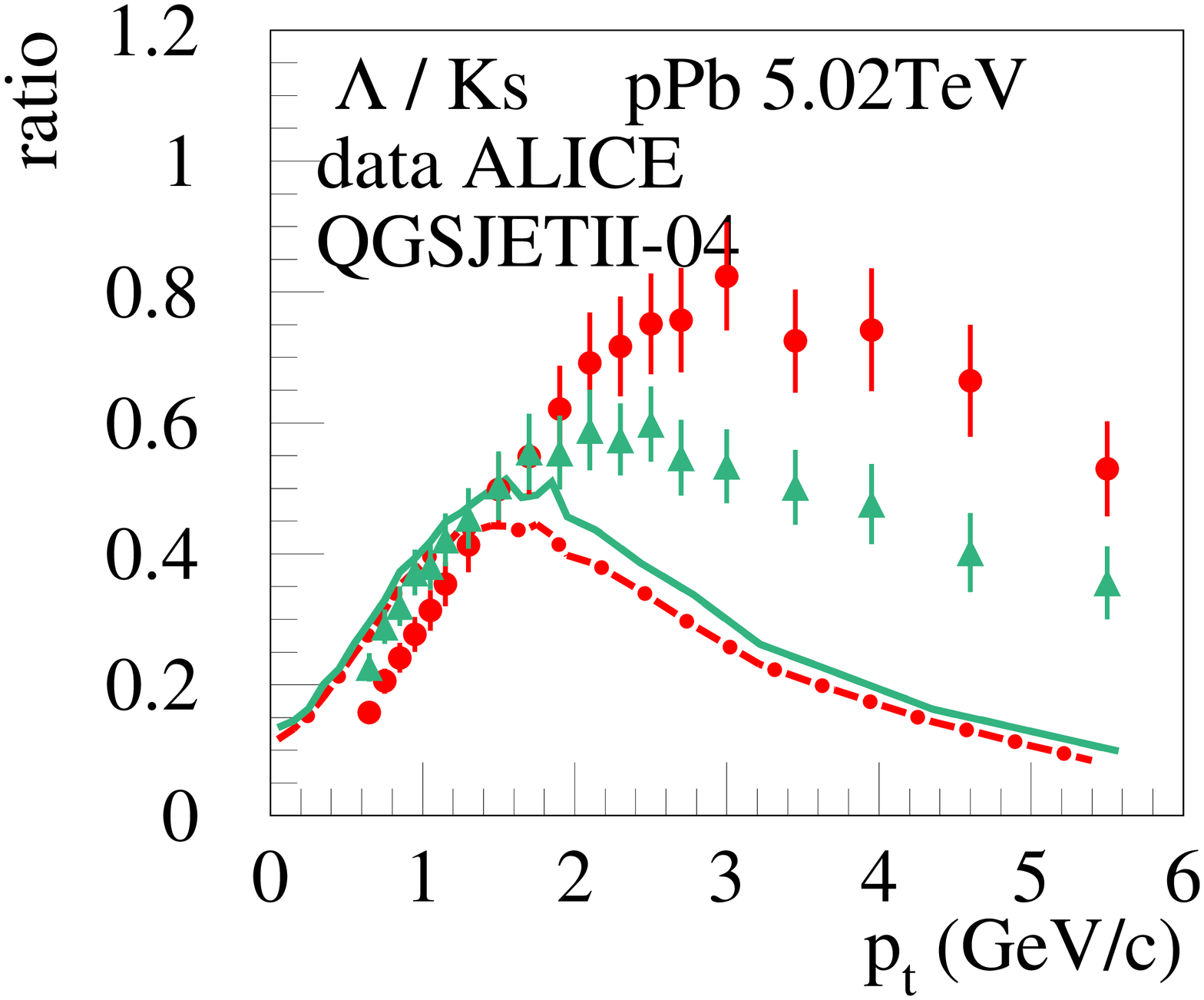}$\qquad$\includegraphics[angle=270,scale=0.28]{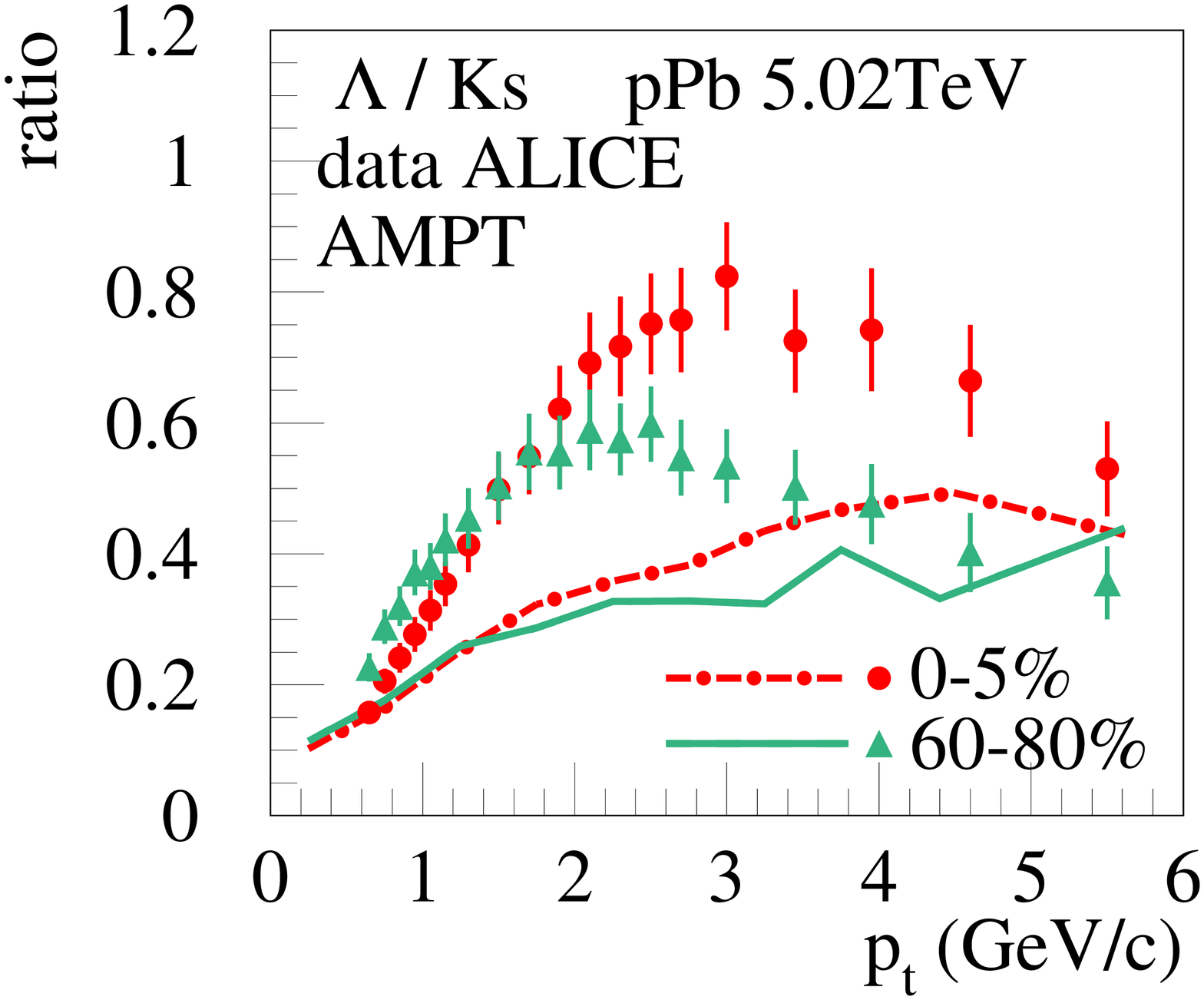}\vspace*{-0.6cm}
\par\end{center}

\begin{center}
\includegraphics[angle=270,scale=0.28]{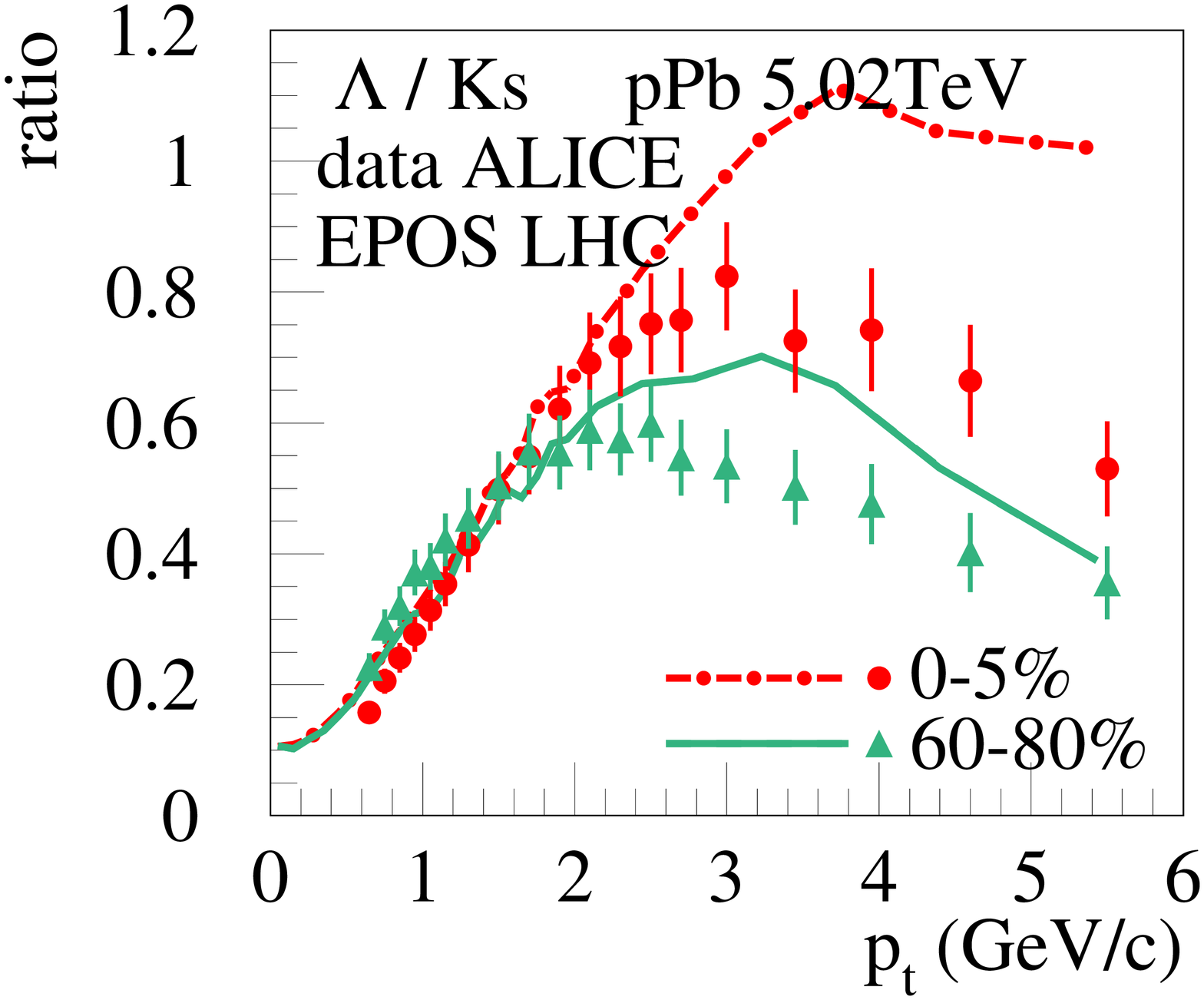}$\qquad$\includegraphics[angle=270,scale=0.28]{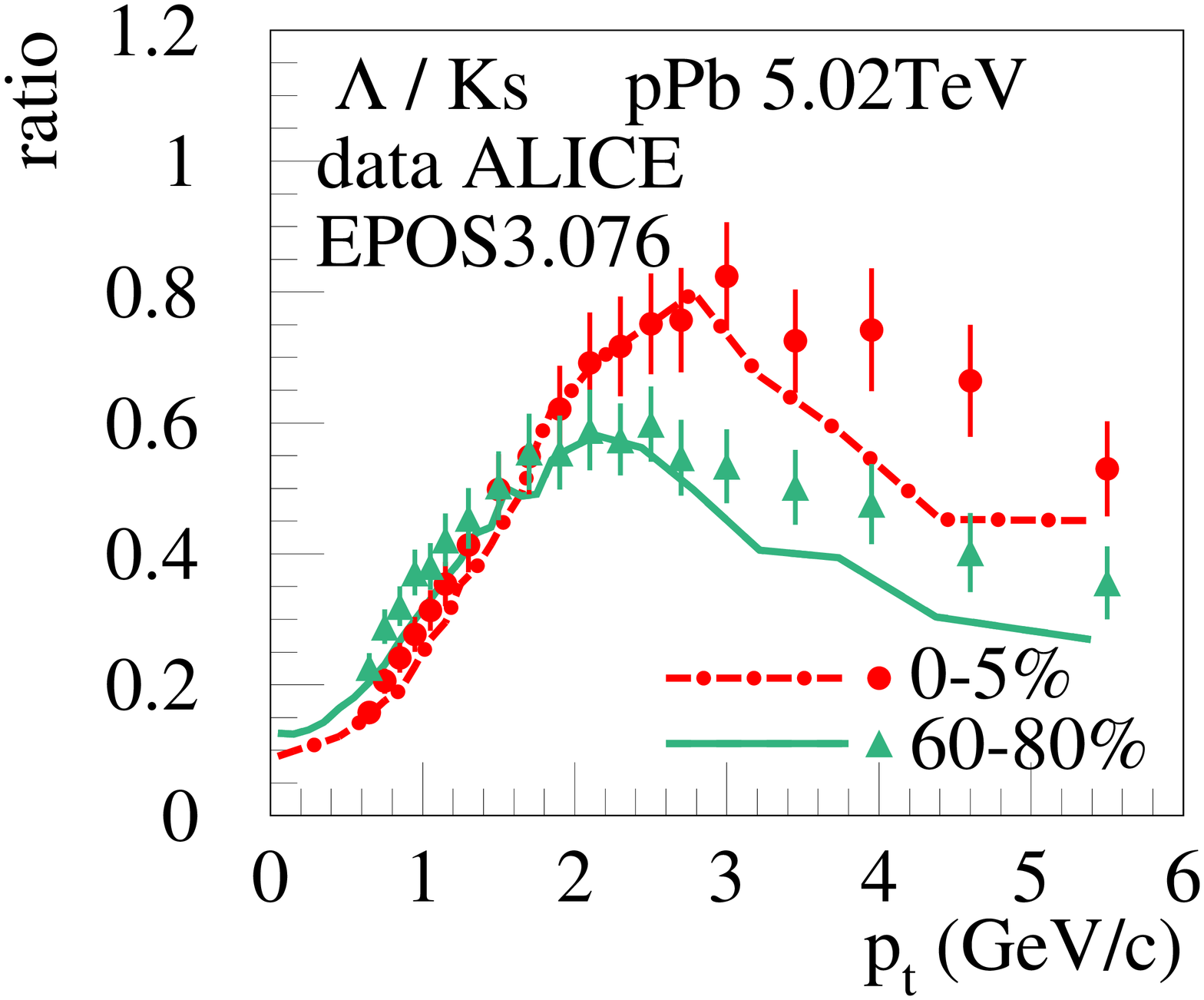}
\par\end{center}%
\end{minipage}
\par\end{raggedleft}

\noindent \caption{(Color online) $\Lambda$ over $K_{s}$ ratio as a function of transverse
momentum in p-Pb scattering at 5.02 TeV, for the 0-5\% highest multiplicity
(red dashed-dotted lines, circles) and 60-80\% (green solid lines,
triangles).\label{fig:selid07} \protect \\
~}
\end{figure}
The experimental shapes of the $p_{t}$ spectra change considerably,
getting much harder with increasing multiplicity. In QGSJETII, having
no flow, the curves for the different multiplicities are identical.
The AMPT model shows some (but too little) change with multiplicity.
EPOS$\,$LHC goes into the right direction, whereas EPOS3 gives a
reasonable description of the data. 

Also ALICE \citet{alice} has measured identified particle production
for different multiplicities in p-Pb scattering at 5.02 TeV. Here,
multiplicity counts the number of charged particles in the range $2.8<\eta_{\mathrm{lab}}<5.1$.
It is useful to study the multiplicity dependence, best done by looking
at ratios. In fig. \ref{fig:selid07}, we show the lambda over kaon
($\Lambda/K_{s}$) ratios. Here, a wide transverse momentum range
is considered, showing a clear peak structure with a maximum around
2-3 GeV/c, with a more pronounced peak for the higher multiplicities.
QGSJETII and AMPT cannot (even qualitatively) reproduce these structures.
EPOS$\,$LHC shows the right trend, but the peak is much too high
for the high multiplicities. EPOS3 is close to the data. Flow seems
to help considerably!

\begin{figure}[tb]
\begin{centering}
\includegraphics[angle=270,scale=0.24]{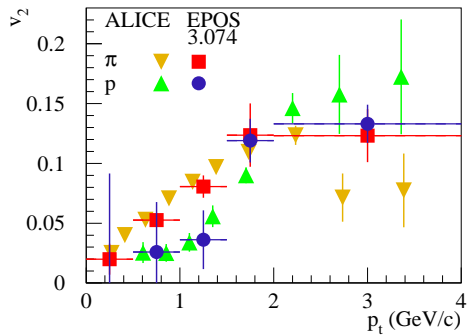}
\par\end{centering}

\caption{(Color online) Elliptical flow coefficients $v_{2}$ for pions and
protons. We show ALICE results (triangles) and EPOS3 simulations (squares
and circles). Pions appear red and yellow, protons blue and green.
\label{fig:v2}}
\end{figure}

Finally, we sketch very briefly results on elliptical flow $v_{2}$
obtained from dihadron correlations, showing ALICE results \citet{alice13b}
and EPOS3 simulations, see ref. \citet{epos3v2} for details. In fig.
\ref{fig:v2}. we plot %
$v_{2}$ as a function of $p_{t}$. Clearly visible in data and in
the simulations: a separation of the results for the two hadron species:
in the $p_{t}$ range of 1-1.5 GeV/c, the proton result is clearly
below the pion one. Within our fluid dynamical approach, the above
results are nothing but a {}``mass splitting''. The effect is based
on an asymmetric (mainly elliptical) flow, which translates into the
corresponding azimuthal asymmetry for particle spectra. Since a given
velocity translates into momentum as $m_{A}\gamma v$, with $m_{A}$
being the mass of hadron type $A$, flow effects show up at higher
values of $p_{t}$ for higher mass particles.\\

To summarize : Comparing experimental data on identified particle
production to EPOS3 and other Monte Carlo generators, we conclude
that hydrodynamical flow seems to play an important role not only
in A-A scattering, but also in p-A and (not shown here) in p-p, supporting
the hypothesis of a unified description.\\

This research was carried out within the scope of the GDRE (European
Research Group) {}``Heavy ions at ultrarelativistic energies''.
Iu.K acknowledges support by the National Academy of Sciences of Ukraine
(Agreement 2014) and by the State Fund for Fundamental Researches
of Ukraine (Agreement 2014). Iu.K. acknowledges the financial support
by the LOEWE initiative of the State of Hesse and Helmholtz International
Center for FAIR. B.G. acknowledges the financial support by the TOGETHER
project of the Region of {}``Pays de la Loire''.

~

\end{document}